\documentclass[a4paper]{jpconf}
\usepackage{graphicx}
\begin{document}
\title{Ground-Based Gamma-Ray Astronomy at Energies Above 10~TeV: Searching for Galactic PeV Cosmic-Ray Accelerators}

\author{Gavin Rowell$^1$,  Felix Aharonian$^1$, Alexander Plyasheshnikov$^{1,2}$}

\address{1. Max Planck Institut f\"ur Kernphysik, Heidelberg D-69029 Germany\\
  2. Altai State University, Barnaul, Russia}

\ead{gavin.rowell@mpi-hd.mpg.de}

\begin{abstract} 
 The origin of Galactic CRs up the knee energy remains unanswered and provides strong motivation for the study of $\gamma$-ray sources
 at energies above 10~TeV. We discuss recent results from ground-based $\gamma$-ray Cherenkov imaging systems at these energies as well as
 future observational efforts in this direction. The exciting results of H.E.S.S. give clues as to the nature of Galactic CR accelerators,
 and suggest that there is a population of Galactic $\gamma$-ray sources with emission extending beyond 10~TeV. A dedicated system of 
 Cherenkov imaging telescopes optimised for higher energies appears to be a promising way to study the multi-TeV $\gamma$-ray sky.
\end{abstract}

\section{Introduction}
The origin of hadronic Cosmic-Rays (CRs) remains an unsolved problem since their discovery nearly 100 years ago. A major area of focus
concerns the Galactic component of CRs, which is thought to be dominant at least up to the {\em knee} ($\sim10^{15}$~eV; 1 PeV), 
the energy at which a spectral break is seen in the CR spectrum. It is generally believed that shell-type supernova remnants (SNRs) are 
responsible for the acceleration of the Galactic component of CRs based on energy budget considerations (of SNRs and of the Galactic
CRs), and the good agreement between the observed properties of the CR spectrum and diffusive shock acceleration (DSA) theory. While DSA theory is
able to explain many aspects of the observed Galactic CRs, a controversial aspect concerns the maximum particle energies attained
\cite{Kirk:1,Drury:1}.
A maximum energy limit of $\sim10^{14}$~eV is generally predicted \cite{Lagage:1} which falls somewhat short of the knee energy. 
Considerably higher energies approaching the
{\em ankle} ($\sim10^{18}$~eV) are thought possible if one accounts for significant fluctuations in the magnetic field \cite{Bell:1,Lucek:1}.  
Alternative source scenarios have also been considered such as local Gamma-Ray-Bursts (GRB) \cite{Waxman:1,Dermer:1}. 
In the conventional context of SNR acceleration of CRs, so-called Superbubbles, which benefit from the combined effects of many SNR shocks and 
possibly also from the stellar winds of OB associations, may also accelerate CRs to well beyond the knee \cite{Drury:2,Bykov:1,Parizot:1}.

We are therefore left with critical questions concerning the origin of CRs at and above the knee. Unfortunately, identifying localised sources of 
Galactic CRs at such energies is impossible by direct detection since the CR trajectories are 
are randomised by Galactic magnetic fields. However, the interaction of CRs with local matter can produce a 
flux of $\gamma$-rays\footnote{and neutrinos} in the multi-GeV to multi-TeV energy band which act as an accessible {\em tracer} of
CR accelerators. Moreover, the energy spectrum of the $\gamma$-ray flux is expected to follow that of the parent CR spectrum at the
site of $\gamma$-ray production, thus providing important information concerning the CR acceleration properties.
We note that the interaction of CRs with ambient matter proceeds with a typical inelasticity of order $\sim$0.15. A source of $\gamma$-rays, 
when interpreted in the above framework, is therefore a source of hadronic CRs of energies roughly a factor 10 higher. Establishing beyond doubt 
a hadronic origin for a $\gamma$-ray flux
is non-trivial since we must often deal with additional leptonic sources of $\gamma$-ray production such as
inverse-Compton scattering and non-thermal Bremsstrahlung. These latter components are often diminished at higher energies due to the strong cooling 
suffered by electrons in magnetised post-shock environments. Morphology differences between the hadronic and leptonic components are also expected,
due to a number of physics issues such as energy-dependent diffusion. Spatially-resolved spectra and energy-resolved morphology studies 
of $\gamma$-ray sources are necessary to disentangle the hadronic and leptonic components. 
In particular, establishing the spectra and morphology of sources of $\gamma$-rays at energies 10~TeV and above is vital to understanding where 
and how hadronic CR acceleration to the knee is taking place.

The detection of TeV $\gamma$-rays is the exclusive domain of ground-based methods, and a major goal of the field 
is to establish the types of sources responsible for Galactic CRs. Given the maximum energy problem discussed above, 
it is not sufficient however just to establish numerous $\gamma$-ray source types and their spectral slope, but to clearly establish the 
maximum energies of their $\gamma$-ray fluxes well into the multi-TeV domain.
The performance of ground-based $\gamma$-ray detectors at energies $E>10$~TeV will therefore attract increasing interest over the coming years.

\section{Present Ground-Based Results at Energies $E>10$ TeV}

The stereoscopic imaging atmospheric Cherenkov technique is employed by all of the present ground-based detectors such as H.E.S.S., 
CANGAROO-III, VERITAS and MAGIC (stereoscopy will be employed with an additional telescope - MAGIC-II). These systems have built on the 
experience from earlier detectors like the HEGRA IACT-System \cite{HEGRA_Puhl} 
which established stereoscopy, and the Whipple 10 metre telescope \cite{Weekes:1}, which first demonstrated the power of imaging.
Survey instruments such as the particle detector arrays MILAGRO, TIBET, KASCADE and CASA-MIA for example also operate in the TeV to PeV domain.
We will however not cover results from these experiments, a selection of which may be found in \cite{McKay:1,Borione:1,Cui:1,Antoni:1,Atkins:1} 
and references therein.
 
The TeV results from H.E.S.S. (see summary in \cite{Rowell:1}) have signalled a turning point for TeV $\gamma$-ray astronomy in that the number of 
sources has grown rapidly, and individual source morphology is revealed for the first time \cite{HESS_RXJ1713,HESS_VelaJun}. 
For the shell-type SNR RX~J1713.7$-$3946 there are now sufficient photon statistics to permit energy-resolved morphology, and spatially-resolved 
spectral studies \cite{HESS_RXJ1713_paperII}.
Over half of the more than 20 TeV $\gamma$-ray source have a Galactic origin\footnote{This is somewhat biased by the H.E.S.S. observation programme, 
which is split roughly 50:50 between Galactic and extragalactic targets.} \cite{HESS_galscan}.
Many of these new Galactic sources, with integral fluxes in the range few \% to 15\% Crab (at $E>1$~TeV), have plausible associations with nearby 
shell-type SNR or plerionic SNR (pulsar-wind-nebulae). 
A striking observation is that the majority of the new sources exhibit generally hard photon spectra, typically characterised by an unbroken power-law  
$dN/dE \sim E^{-\Gamma}$ with index $\Gamma$ in the range 2.0 to 2.5,
and extending to up to 40~TeV. The majority are also extended in morphology, with diameters
ranging from a few arcmin to a degree. These results are in line with expectations of
DSA theory and give strong hints as to the type of Galactic CR accelerators. 
The establishment of spectra above energies $\sim$10~TeV is often limited however by statistics, given the available observation time and
sensitivity of H.E.S.S.
Deeper observations will in fact be performed on selected H.E.S.S. sources such as RX~J1713.7$-$3946, which will be observed at high zenith angles 
(this method is discussed shortly) during 2005 in a strong effort to establish the spectrum beyond 40~TeV. 

\begin{figure}
\begin{center}
\includegraphics[width=30pc]{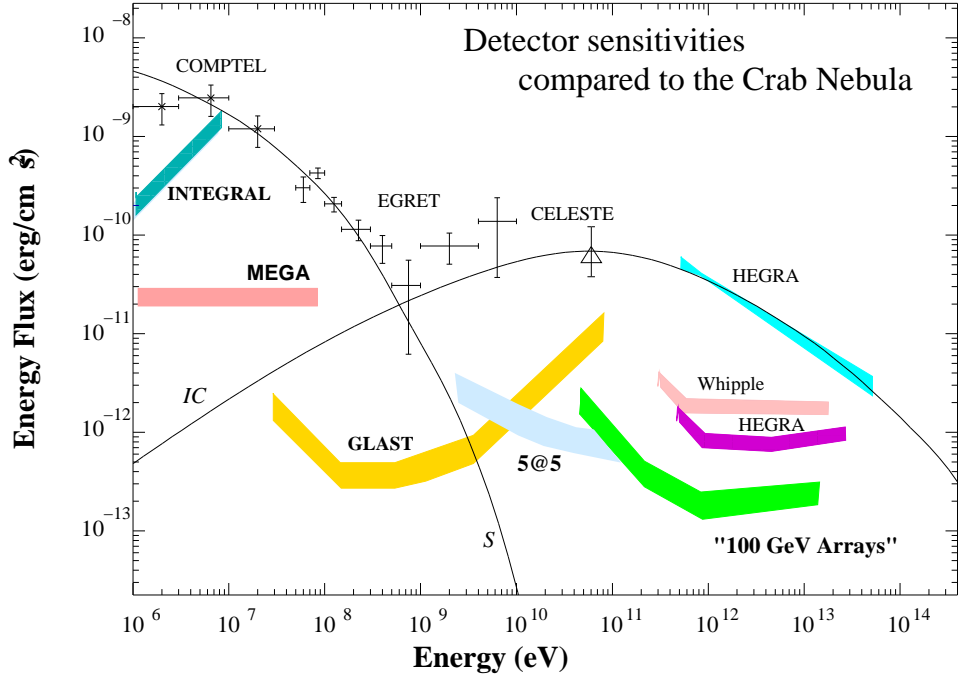}%
\end{center}
\caption{\label{fig:sens} Energy flux sensitivities for various instruments operating in the $\gamma$-ray domain,
                          compared the measured and predicted (synchrotron and inverse-Compton) fluxes from the Crab Nebula.
                          For the ground-based instruments, limited here to Cherenkov imaging systems  
                          ``100-GeV Arrays'' (representing H.E.S.S., VERITAS and CANGAROO-III), HEGRA IACT-System, Whipple, 5@5), 
                          a 50~h 5$\sigma$ detection limit is required. See \cite{MAGIC} for details of the MAGIC sensitivity. 
                          For the space-based detectors (GLAST, INTEGRAL, MEGA) a 10$^6$~sec obs time is specified. The reported $\gamma$-ray
                          fluxes from HEGRA, CELESTE, EGRET and COMPTEL are also shown. From \cite{Felix_book}. See Figure~\ref{fig:crab}
                          for a more recent Crab spectrum from HEGRA.}
\end{figure}

\begin{figure}
\includegraphics[width=20pc]{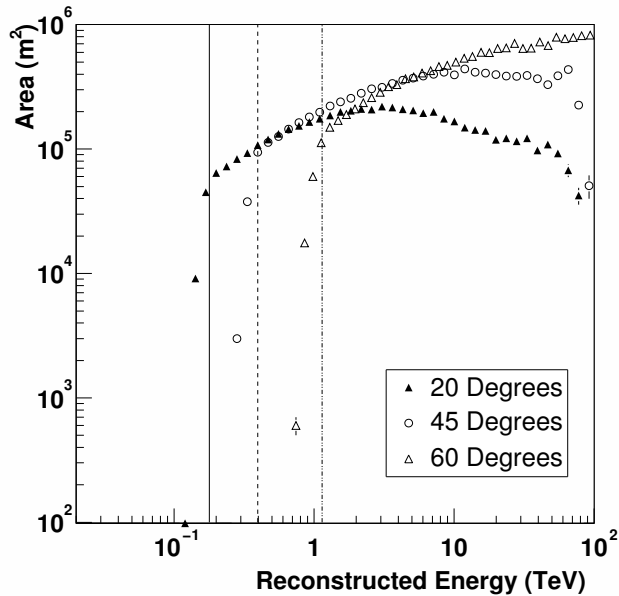}\hspace{2pc}%
\begin{minipage}{15pc}
\vspace{-8cm}
\caption{\label{fig:aeff}H.E.S.S. effective collecting area vs. energy (following standard $\gamma$-ray selection cuts) for a variety of 
      zenith angles (20$^\circ$, 45$^\circ$, 65$^\circ$) \cite{HESS_crab}. Vertical lines indicate the energy threshold (peak differential     
      rate for a Crab-like spectrum) applicable to each
      zenith angle.
      An increase in the effective collecting area at higher energies is clearly noticed for higher zenith angle observations.}
\end{minipage}
\end{figure}

H.E.S.S. is optimised for $\gamma$-ray detection in the energy range $\sim100$~GeV up to a few 10s of TeV. The specification of the four H.E.S.S. 
telescopes, with mirror area $\sim$120~m$^2$, 0.12$^\circ$ camera pixel resolution, 5 deg camera field of view (FoV), 
and a telescope spacing of 120~m, gives H.E.S.S. an optimal sensitivity in the 1 to 10~TeV range,
where the CR background rejection is strongest, and the angular resolution is better than a few arcminutes 
(see sensitivity curve Figure~\ref{fig:sens}). In this energy range an energy flux sensitivity of 
$\sim$10$^{-13}$ erg cm$^{-2}$s$^{-1}$, or 1\% Crab flux, in 25~h observation time at low zenith angles (LZA) is achieved.
The performance at higher energies begins to decrease since the camera FoV, and telescope spacing (defining the physical size of the array) 
imposes a limit on the detection efficiency.
This limitation may be offset somewhat by employing high-zenith angle (HZA) observations \cite{Somers:1}, where the observation zenith 
angle $z$ is larger than $\sim 55^\circ$. At such angles, the longer atmospheric path length means that the
air shower is viewed at greater distances from the telescope. As a result, higher energy EAS preferentially trigger the detector from 
considerably larger distances (impact parameters) than they would at lower zenith angles, thereby increasing the effective collection area for a
given energy, but at the same time increasing the energy threshold. Using this technique, the effective collection area of a ground-based system like  
H.E.S.S. at high energies approaches 1~km$^2$ (Figure~\ref{fig:aeff}). 

\begin{figure}
\includegraphics[width=20pc]{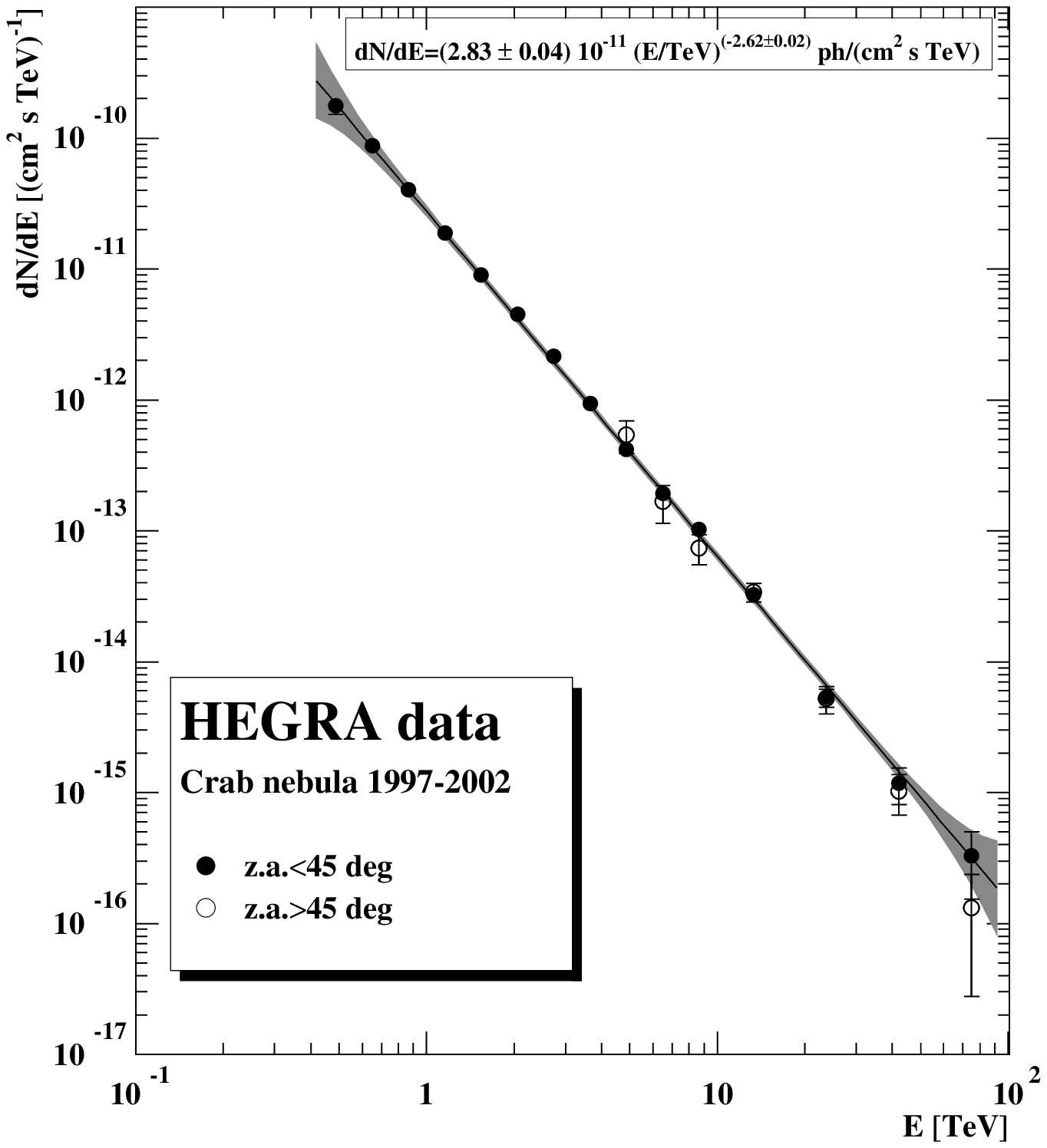}\hspace{2pc}%
\begin{minipage}{15pc}
\vspace{-8cm}
\caption{\label{fig:crab} Energy spectrum of the Crab Nebula from HEGRA IACT-System observations \cite{HEGRA_crab}. Data from low and high zenith
         angles are compared. The spectrum extends to $\sim$70~TeV. The highest energy spectral point has an excess significance $+2.7\sigma$.}  
\end{minipage}
\end{figure}

HZA observations have already been demonstrated effectively in the study of the Crab Nebula by the HEGRA IACT-System \cite{HEGRA_crab} 
and Mkn~421 by H.E.S.S. \cite{HESS_mkn421}. 
The energy spectrum of the Crab is now established up to energies approaching $\sim$70~TeV after very deep observations ($\sim$380~h) 
with the HEGRA IACT-System (Fig.~\ref{fig:crab}).
The mechanism put forward to explain the (unpulsed) TeV emission of the Crab centres on the inverse-Compton upscattering of soft photon fields by
accelerated electrons. The inverse-Compton $\gamma$-rays seen up to 70~TeV, along with the  synchrotron emission seen up to GeV energies by EGRET
strongly imply the presence of electron acceleration to PeV energies \cite{HEGRA_crab}.
Thus, the Crab is likely the first source established as a cosmic {\em Pevatron}. It therefore remains to be seen if it, and other Galactic 
sources are capable of accelerating hadrons to similar energies.
H.E.S.S. observations of Mkn~421 were triggered by a flaring state reaching fluxes of up to $\sim$6~Crab. At the H.E.S.S. site
this northern source is visible at zenith angles $\sim65^\circ$, yielding an energy threshold about 2~TeV, however with
collection areas exceeding 1~km$^2$. Despite the fact that
the energy spectrum of Mkn~421 exhibits an exponential cutoff ($dN/dE \sim E^{-\Gamma} \exp(E/E_c)$ where $E_c=3$~TeV) 
limiting the photon statistics above 10~TeV, HZA observations permitted a detection rate of up to 8 $\gamma$-rays per minute.
For the first time, variability studies on time scales of hours were performed on this source in the multi-TeV energy range.  
The HZA observation method has limitations however. Observability issues come into play as most sources do not spend much time at high zenith
angles. Indeed this method is better suited, for instruments in the south, to viewing sources with northern declinations in order to maximise
the available observation time per night (the opposite of course is true for southern sources viewed from the north). Other issues such
as atmospheric stability can also reduce the overall efficiency of the observations.

\section{A Dedicated Ground-Based Instrument for Energies $E>10$~TeV}

It is clear that we are only just beginning to sample the $\gamma$-ray sky at energies $\sim 10$ to 100~TeV at 1\%~Crab flux sensitivties.
Over the past decade, is has become apparent that the stereoscopic atmospheric Cherenkov imaging technique has proved highly 
successful for $\gamma$-ray detection in the 100~GeV to $\sim$10~TeV energy range. It is therefore worthwile investigating the same technique
optimised for higher energies. The Monte-Carlo simulations of Plyasheshnikov et~al. \cite{Plya:1} first looked at the potential of 
the stereoscopic atmospheric Chereknov imaging technique for $E>10$~TeV $\gamma$-ray astronomy, and we draw on those results here.
They considered a {\em cell} of four imaging Cherenkov telescopes, equipped with focal plane cameras
of photomultiplier's (PMTs) of various sizes and telescope spacing. The main motivation was to investigate the potential of the technique,
when applying generally conservative specifications to the telescope design. In this sense a quite conservative quantum efficiency (QE)
(photons to photoelectrons) value of 0.1 was assumed. Monte-Carlo simulations of both $\gamma$-rays and protons viewed at the vertical were
used. Two different pixel sizes for the focal plane detectors (or cameras) 
were considered, 0.3$^\circ$ and 0.5$^\circ$, 
and for each case, a two-level tailcut scheme to remove pixels dominated by skynoise was also applied\footnote{This is also referred to as image 
{\em cleaning}. In H.E.S.S. data analysis, tailcuts of 5 and 10 p.e. are often used.}. A telescope trigger was achieved when at least 2 two adjacent 
pixels meet a defined threshold (10 to 20 p.e.). A stereoscopic trigger was demanded when at least two telescopes have triggered from the same
EAS. In order to gauge the performance 
characteristics of the {\em cell}, a moment-based image parametrisation scheme according to Hillas was applied. This employed the {\em width} parameter
and its stereoscopic counterpart {\em mean-scaled-width} $W$ \cite{Aharon_study} as the image {\em shape} basis for CR background rejection, 
while the image major-axes were used in the reconstruction of arrival directions. A point spread function (PSF) of size
$\sim$0.15$^\circ$ to 0.1$^\circ$ (std. deviation of a fitted Gaussian) was indicated for the {\em cell} when assuming a pixel spacing of 0.3$^\circ$. 
These trigger conditions and analysis algorithms applied to the
{\em cell} were based on the methods used in the HEGRA IACT-System, and now, also employed by H.E.S.S.

\begin{figure}
\includegraphics[width=22pc]{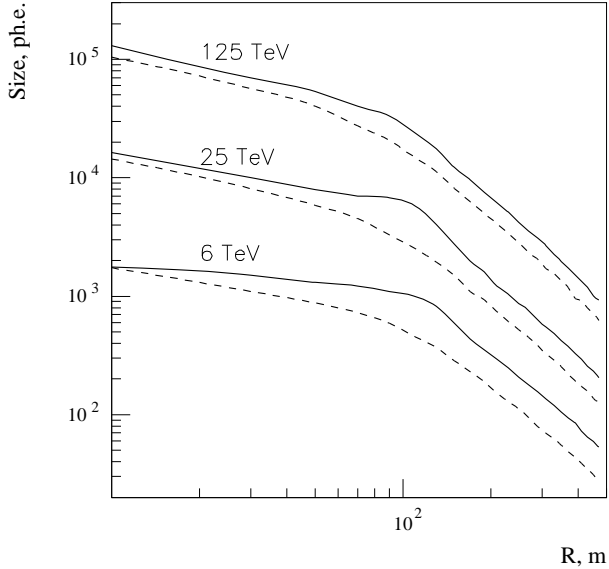}\hspace{1pc}%
\begin{minipage}{15pc}
\vspace{-8cm}
\caption{\label{fig:phot_yield} Average photoelectron yield for Cherenkov images vs. impact parameter ($R$ = distance to the air shower core)
   for a range of energies. Solid curves ($\gamma$-rays); dashed curves (CR protons). A field of view of infinite diameter is used, to accept
   photons from all directions. From \cite{Plya:1}.}  
\end{minipage}
\end{figure}

\begin{figure}[h]
\begin{minipage}{18pc}
\includegraphics[width=20pc]{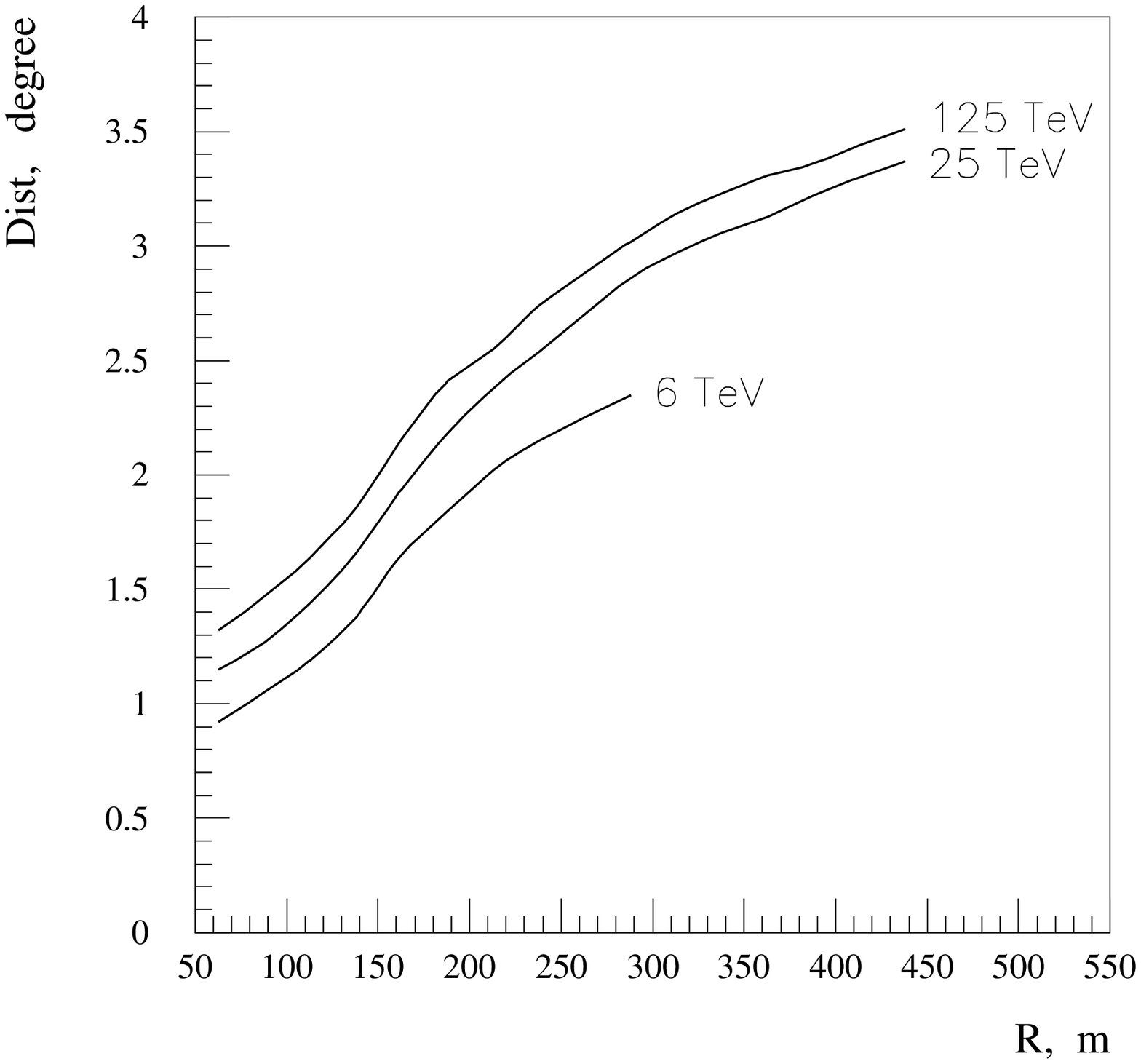}
\caption{\label{fig:dist} Average value of {\em dist} (the image centroid distance from the focal plane centre) vs. impact parameter $R$.
          Images are viewed and reconstructed using a large FoV camera (9.3$^\circ$, with pixel sizes $0.3^\circ$) and pixel trigger threshold
          of 15 p.e. From \cite{Plya:1}.}
\end{minipage}\hspace{2pc}%
\begin{minipage}{18pc}
\includegraphics[width=20pc]{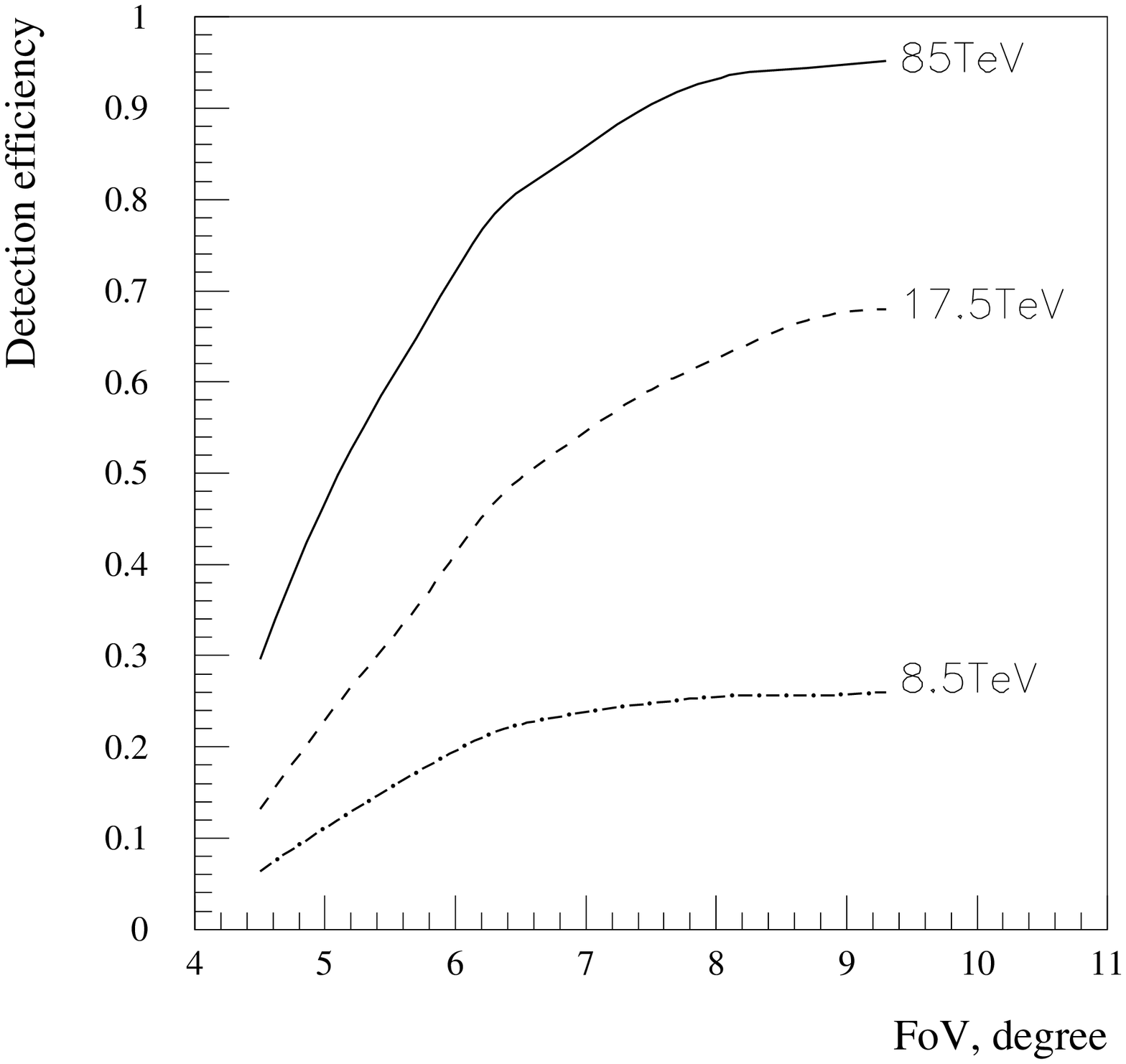}
\caption{\label{fig:spacing} The efficiency (fraction of events triggering relative to those incident) of event trigger vs. the detector 
          FoV for several $\gamma$-ray energies. A telescope spacing of
         500~m is used and pixel trigger threshold of 10 p.e. From \cite{Plya:1}.} 
\end{minipage} 
\end{figure}

It is worth highlighting here the major conclusions of the simulation study:
\begin{itemize}
 \item {\bf Telescope Mirror Area:} For energies $E>10$~TeV, only a modest mirror area $A\sim10$~m$^2$ is required to detect sufficient
Cherenkov photons (or equivalent photoelectrons). Figure~\ref{fig:phot_yield} depicts for both $\gamma$-ray
and CR proton induced air-showers, the number of Cherenkov photoelectrons (p.e.) contained within
an image vs. impact parameter (the distance of the air-shower core to the telescope) and energy.
Experience has shown that $\sim$50 to 100 p.e. per image are required to accurately reconstruct individual image parameters such as {\em width}, 
{\em length} and orientation. A mirror area of $10$~m$^2$ is therefore more than sufficient to detect and reconstruct events with energy $E>10$~TeV
out to impact parameters 500~m or more. 
\item {\bf Focal Plane Camera FoV:} EAS that trigger a telescope out to 500~m will be imaged with considerable angular displacement
in the focal plane, especially when observed at low zenith angles (0$^\circ$ to 45$^\circ$). Figure~\ref{fig:dist} shows 
the relationship of the image centroid distance from the focal plane centre (the so-called {\em dist}
parameter) vs. impact parameter. Images with impact parameters approaching 500~m will trigger up to 4$^\circ$ off-axis.
When one allows for edge effects of the finite FoV, a typical image {\em length} of order 0.8$^\circ$ and the fact that many sources might be extended, 
fields of view in excess of 8$^\circ$ diameter would be considered necessary. Depending on the pixel sizes (0.3$^\circ$ or 0.5$^\circ$), 
one would require between 300 to 800 pixels in order to achieve such fields of view. 
\item {\bf Telescope Spacing:} The efficiency (the fraction of events triggering relative to those incident) of the stereoscopic
trigger depends on the distance between telescopes, on the telescope FoV, and energy of the EAS. The results of Figure~\ref{fig:dist} suggest 
that a large telescope spacing of order 500~m is suited to multi-TeV energies.  
Figure~\ref{fig:spacing} depicts the detection efficiency for  
four telescopes with spacing 500~m. In fact a spacing in the range 300 to 500~m was found necessary to achieve an 80\% efficiency over the 
desired energy range, when employed in combination with very large FoV cameras.
\end{itemize}

Assuming then a telescope spacing of 500~m, an energy threshold (the peak differential rate) of 7 to 10~TeV was achieved for a
point source with a Crab-like energy spectrum.
For a source with a Crab-like flux, detection rates of 3.5 $\gamma$-rays and $\sim$700 CR background events per hour (over the entire FoV for
the CR background) were achieved. A combination of a cut on the angular distance 
$\theta<0.1^\circ$ between the reconstructed and assumed directions in addition to a cut on the {\em scaled-width} parameter (eg. $W<1.0$) 
accepted $<$0.03\% of CR background events, whilst retaining 50\% of $\gamma$-ray events. After these cuts, a Crab-like source yields
detection rates of 1.7 $\gamma$-rays per hour against a CR background rate of 0.2 events per hour, giving a signal to noise ratio approaching
a factor 10 and excess significance $\sim 2\sigma$/$\sqrt{\rm hr}$\footnote{Here we assume that the CR background is estimated from a region in the FoV
with solid angle a factor 10 larger than that of the source. Eq. 17 from Li \& Ma \cite{Li:1} with $\alpha=0.1$ is then used to estimate the 
significance.}.  It is interesting to compare these numbers with the detection rate obtained on real Crab data, for energies $E>7$ to 10~TeV from 
both the HEGRA IACT-System and H.E.S.S. detectors. For HEGRA, a detection rate of $\sim$2 $\gamma$-rays and 0.3 CR per hour is achieved at 
$\sim 2\sigma$/$\sqrt{\rm hr}$ \cite{HEGRA_crab}. 
For H.E.S.S., the Crab culminates at middle-range zenith angles of $\sim 45^\circ$, improving the collecting area somewhat at higher energies
(Figure~\ref{fig:aeff}). On the Crab, H.E.S.S. achieves a $\gamma$-ray rate near culmination of $\sim6$ $\gamma$-rays and 0.4 CRs per hour 
at $\sim 5\sigma$/$\sqrt{\rm hr}$. 
If one factors in the improved collection area for the H.E.S.S. Crab result, the performance of a single {\em cell} of small telescopes is 
quite similar to that of H.E.S.S. and the HEGRA IACT-System at the high energies considered. 
Another important performance measure is the energy resolution, which is estimated at $\sim$25\%. 
This is sufficient to reconstruct energy spectra accurately and in particular determine cutoff features. 

Of course the majority of sources under future study will emit at sub-Crab flux levels. The H.E.S.S. results suggesting a population of sources in the 
5 to 15\% Crab flux range ($E>1$~TeV),  would demand considerable observation time from a single {\em cell}. This is helped somewhat by the hard
spectra of the new sources (which would have fluxes $\sim 0.2$~Crab-flux level for $E>10$~TeV).
In order to detect sources with high statistics in reasonable time ($<100$~h), \cite{Plya:1} also suggest the use of multiple {\em cells}
of 4-telescope arrays. For example they estimate a total of 25 telescopes (arranged as 16 {\em cells} of 4-telescopes) would achieve detection
rates (after cuts) on the Crab of $\sim 1250$ $\gamma$-rays and 150 CRs in 50~h observation. This observation time would be sufficient for 
extraction of a spectrum for a 0.1~Crab-flux source. 

It should again be stressed that the performance of a high energy array of telescopes would be improved for hard-spectra sources
(the simulation study assumed a softer Crab-like spectrum), like those
being uncovered by H.E.S.S., and indeed those expected to be the sources of hadronic CRs. Finally we note that the detectability of $\gamma$-ray sources   
will be affected by the attenuation of $\gamma$-rays due to pair production on the microwave background \cite{Coppi:1}. This loss strongly peaks 
at $\sim$1~PeV with a mean free path of $\sim$10~kpc, meaning that PeV $\gamma$-ray astronomy would be limited to searching for relatively 
nearby Galactic sources.

\section{Conclusions}

Studies of $\gamma$-ray sources at energies above 10~TeV are vital if we are to understand how and where Galactic cosmic-rays are accelerated
to  and above the knee. Since the present ground-based telescopes such as H.E.S.S. are optimised for energies $E\leq$10~TeV, it is suggested that
a new dedicated system of telescopes be employed to open up the $\gamma$-ray sky at higher energies. Monte-Carlo simulations
of a {\em cell} of four modest-size telescopes employing the same stereoscopic imaging technique as in H.E.S.S. and the HEGRA IACT-System
suggest that sufficient sensitivity is achieved in the $E>10$~TeV range. These encouraging results, from rather conservative
simulations, suggest that the stereoscopic imaging approach is a viable method to explore the multi-TeV sky. More detailed simulations
are now underway.

\end{document}